# Molecular dynamic simulation on the density of titanium dioxide and silver water-based nanofluids

M. M. Heyhat[*1], M. Abbasi[1], A. Rajabpour[2,3]

[1*] Faculty of Mechanical Engineering, Tarbiat Modares University, Tehran, Iran, mmheyhat@modares.ac.ir

[2] Mechanical Engineering Department, Imam Khomeini International University, Qazvin, Iran, rajabpour@eng.ikiu.ac.ir

[3] School of Nano Science, Institute for Research in Fundamental Sciences (IPM), P.O. Box 19395-5531, Tehran, Iran

**Abstract:**

This study utilizes molecular dynamics simulations to scrutinize the influence of type and volume fraction of nanoparticle on the density of nanofluids, i.e. silver (as a hydrophilic case) and titanium dioxide (as a hydrophobic case) water-based nanofluids. Ionic structure of $TiO_2$ distinguishes the solid-liquid interactions with those in silver-water nanofluid. The thickness of nanolayer is estimated from density profile formed around the nanoparticle and is included in the density calculation. It is found that the type of interatomic potential between the nanoparticle and surrounding water can highly affect the thickness of nanolayer. The hydrophobicity/hydrophilicity of nanoparticle is of great importance in nanolayer density and its thickness. It is observed that the difference between the traditional binary mixture and the ternary mixture formula for calculating the density of nanofluid increases by increasing the volume fraction of nanoparticle. This indicates that the effect of nanolayer gets more considerable at high volume fractions.

# Nomenclature

| | | | |
|---|---|---|---|
| d | Nanoparticle diameter (nm) | i | initial state, before mixing |
| E | Energy (J) | nf | nanofluid |
| F | Force (N) | nl | liquid nanolayer |
| m | Mass (kg) | np | nanoparticle |
| M | Molecular mass (gr/mol) | | |
| p | Momentum (kg.m/s) | **Greek letters** | |
| q | Partial charge ratio | $\varphi$ | Volume Fraction |
| r | Distance (m) | $\varepsilon$ | Potential well depth |
| t | Nanolayer thickness (nm) | $\sigma$ | Distance at which the intermolecular potential energy between the two particles is zero |
| T | Temperature (K) | $\rho$ | Density (kg/m$^3$) |
| U | Potential energy (J) | | |
| V | Volume (m$^3$) | **Constants** | |
| | | $k_B$ | Boltzmann constant |
| **Subscript** | | $K_c$ | Electrostatic constant |
| bf | base fluid | $N_A$ | Avogadro constant |
| f | final state, after mixing | | |

## 1- Introduction

The use of colloids of nanoparticles called nanofluid has been grown in various fields of science and engineering including oil recovery [1], $CO_2$ capturing [2], solar energy [3], HVAC systems [4], electronics cooling [5], and many others. Understanding the properties of nanofluids must be accomplished before their introduction in commercial applications. The properties of nanofluids hinge on different factors such as type of nanoparticles, geometry and dimension of nanoparticles, base fluid properties, and nanoparticle concentration [6].

The density of nanofluids is a major determinant of nanofluid flow and heat transfer processes that directly affect the Reynolds number, Nusselt number, friction coefficient and pressure loss [7]. For example, changes of density due to changes in temperature can cause the natural convection. Moreover, density has an important impression in characterizing some properties like thermal diffusivity. Therefore, due to its importance, there are several research works dedicated to study the effective nanofluid density [8–11]. Mahian et al. [12] performed an experimental investigation to assess the stability of zinc oxide nanoparticles in a mixture of ethylene glycol–water. They gauged the density of nanofluids in volume fraction ranges and temperatures of 0-4% and 25–40 °C, respectively. It was observed that by increasing the nanoparticle volume fraction the density of nanofluid augmented noticeably while it decreased slightly with the temperature. They compared their experimental data with theoretical data based on mixture theory [13] and found that the theory underestimated the experimental data. Teng and Hung [9] analyzed the density of $Al_2O_3$/water nanofluid experimentally. They reported the density of nanofluids for different temperatures in the range of 10–40°C. Their outcomes also revealed that the experimental data was not compatible with theoretical values based on the concept of mixing theory for ideal gas mixtures. Moreover, the deviation between theoretical values and experimental data became greater as the nanoparticle concentration increased. They attributed this deviation to interfacial nanolayer, specific surface area and adsorption. By dispersing nanoparticles in a liquid, a more structured but thin layer of liquid is formed at the liquid/solid interface which is called the nanolayer [14].

Sharifpur et al. [15] examined the density of four various nanofluids, i.e. $SiO_2$-water, MgO-glycerol, $SiO_x$-EG 60%-water 40%, and CuO glycerol experimentally. They introduced a new model which considered the nanolayer influence on density of nanofluid. Their results indicated that the nanolayer had a significant role in nanofluid density, especially at higher volume concentration of nanoparticles. Heyhat et al. [16] scrutinized the effect of nanolayer on nanofluid density by using molecular dynamics (MD) simulations. Their results indicated that the nanolayer formation led to base fluid contraction. They introduced a novel ternary mixture density model considered the nanofluid as a three- ingredient mixture consisted of base fluid after blending with nanoparticles, nanoparticle and nanolayer. The outcomes indicated the significant role of nanolayer on nanofluid density. Recently, Abbasi et al. [17] conducted classical MD simulations to investigate the effect of particle shape and type of base fluid on density of nanofluid based on the ternary mixture formula. They concluded that planar nanoparticle has the highest density of nanolayer among planar, spherical and rod-shaped nanoparticles, most probably due to having flat surfaces with no curvature. Moreover, Mortazavi et al. [18] showed that it was possible to examine the properties of complex microstructures without prior knowledge of building blocks by using machine learning interatomic potentials as an alternative to classical MD simulations.

Followed by the previously published researches [16–17], the influences of type and volume fraction of nanoparticle on the density of nanofluids using ternary mixture model are examined in this paper. In this regard, molecular dynamic simulations of silver (a hydrophilic surface) and $TiO_2$ (a hydrophobic surface) nanoparticles suspended in water are carried out at different nanoparticle volume fractions. The construction of ordered nanolayer is analyzed and then by calculating the nanofluids density the influence of this layer is surveyed.

**2- Simulation Method and Models**

The volume percentages studied in this paper are 2, 4 and 6%, and the shape of the nanoparticles are spherical. The TIP4P/2005 [19] force-field was used to calculate the intermolecular forces between water particles. In this potential, the partial charge of hydrogen is equal to 0.5564 e and the partial charge of oxygen is equal to -1.1128 e, where e is the unit of charge of an electron. The angle between the bonds of a water molecule is 104.52° and the length of the oxygen hydrogen bond is 0.9572 Angstrom. The interaction between hydrogen atoms is neglected because of their little mass. To calculate the intermolecular forces between water particles, the Lennard-Jones (LJ) potential function plus a Columbic term is used:

$$E_{ab} = \sum_{i}^{a}\sum_{j}^{b} \frac{K_c q_i q_j}{\mathbf{r}_{ij}} + 4\varepsilon_{ij}\left[\left(\frac{\sigma_{ij}}{r_{ij}}\right)^{12} - \left(\frac{\sigma_{ij}}{r_{ij}}\right)^{6}\right] \quad (1)$$

where $a$ and $b$ refer to two water molecules, $i$ and $j$ are related to the atoms in a molecule, $K_C$ shows an electrostatic constant, $\varepsilon$ is the potential well depth and $\sigma$ is the distance at which the Lennard-Jones potential energy between the two molecules is zero.

In order to investigate the role of nanoparticle material in the ternary mixture model for density of nanofluid, $TiO_2$ and silver nanoparticles are simulated. The ionic structure of $TiO_2$ nanoparticle distinguishes it from atomic structure of silver. Three crystalline phases of $TiO_2$ nanoparticles have been identified so far, including rutile, anatase and brookite. These three crystalline phases differ in their spatial group, crystal dimensions, and crystal structure. The thermodynamic stability of these three phases is also different and is largely dependent on the size of the particle. At small sizes (for example, at a particle diameter of less than 14 nm), anatase is more stable than rutile [20].

Table 1 shows the experimental parameters of anatase phase of $TiO_2$ nanoparticle for simulation in Materials Studio software [21].

Table 1. Unit cell parameters taken from experimental results for $TiO_2$ in the anatase phase [22]

| crystal structure | a (Å) | b (Å) | c (Å) |
|---|---|---|---|
| Tetragonal | 3.785 | 3.785 | 9.513 |

A survey of the literatures [23–26] has indicated that the Matsui-Akaogi force-field [27] is widely used to simulate $TiO_2$ nanoparticle. The atomic interactions between the titanium atom and the oxygen atom in TiO$_2$ are determined by the electrostatic force, which is the partial charge of the titanium atom equal to 2.19 e and the partial charge of the oxygen atom equal to 1.908 e. Van der Waals forces between the particles are also calculated by Buckingham's potential. The general form of the Matsui-Akaogi force-field is as follow:

$$U(r_{ij}) = A_{ij} exp\left(-\frac{r_{ij}}{\rho_{ij}}\right) - \frac{C_{ij}}{r_{ij}^6} + \frac{q_i q_j}{r_{ij}} \quad (2)$$

where $r_{ij}$ is the distance between the two atoms $i$ and $j$, and $q_i$ is the electric charge of atom $i$. The coefficients of Buckingham potential in the Matsui-Akaogi force-field are presented in Table 2.

Table 2. Coefficients of Matsui-Akaogi force-field in simulation of TiO$_2$ nanoparticle [27]

| Interaction type | $A_{ij}$ (kcal/mol) | $C_{ij}$ (kcal/mol$^{-1}$Å$^6$) | $\rho_{ij}$ (Å) |
|---|---|---|---|
| Titanium -Titanium | 717654 | 121.07 | 0.154 |
| Titanium - Oxygen | 391053 | 290.40 | 0.194 |
| Oxygen - Oxygen | 271719 | 696.94 | 0.234 |

To validate the TiO$_2$ model, it was simulated separately at 30°C without the presence of water and its bulk density was calculated. The density value was 3.91 g/cm$^3$, which is in good agreement with experiment [28].

The interactions between the nanoparticle and water are expressed by the Lorentz-Berthelot combining rules as follows.

$$\sigma_{ij} = \frac{(\sigma_{ii} + \sigma_{jj})}{2} \quad (3)$$

$$\varepsilon_{ij} = \sqrt{\varepsilon_{ii} \cdot \varepsilon_{jj}} \quad (4)$$

The Lennard-Jones and Columbic potential have been used for interactions between water and TiO$_2$, which is verified by Park and Aluru [29] who calculated contact angle of the water droplet on the $TiO_2$ surface using MD simulation. The LJ parameters used in this study are tabulated in Table 3.

Table 3. LJ potential coefficients in the current work [29–30]

| Interaction | σ (Å) | ε (kcal/mol) |
|---|---|---|
| Ag–Ag | 2.64 | 7.95 |
| $H_2O$–$H_2O$ | 3.16 | 0.182 |
| Ag–$H_2O$ | 2.9 | 1.21 |
| Ti–Ow | 2.99 | 0.056 |
| O–Ow | 3.54 | 0.097 |
| Ti–Hw | - | 0 |
| O–Hw | - | 0 |

Ow: O in water
Hw: H in water

Nanoparticles with diameter of 4 nm consist of 1985 Ag atoms or 2940 Ti/O atoms (980 Ti and 1960 O) surrounded by 53784 water molecules in a three-dimensional box with 12nm side length. Thus, the volume concentration of nanoparticles is around 2 percent. A few numbers of atoms have been removed from the surface of the $TiO_2$ nanoparticle such that the nanoparticle is electrically neutral.

The boundary conditions considered in the X, Y, and Z directions are periodic. To prevent the particle from translational movement inside the simulation box, the center of the nanoparticle is connected to a spring with a constant of 100 kcal.mol$^{-1}$Å$^{-2}$. Figure 1 schematically shows the water simulation box and the nanoparticle. All molecular dynamics simulations are performed by LAMMPS [31] package.

The electrostatic and short-range forces are computed using the Ewald summation technique and the cut-off method, respectively. The time step considered in all simulation steps is 1 fs. The simulation steps are as follows: i) NVT run for 20 ps using Nosé-Hoover thermostat, ii) NPT run for 20 ps using Nosé-Hoover barostat, iii) and then the final NVE run for 40 ps to record data and to calculate the desired quantities.

In NVT and NPT runs, the temperature and the pressure are controlled at 298.15 K and 1 atm, respectively. The Packmol [32] package was employed to generate the initial atomic positions.

The LAMMPS input script used in the present work has already been used in Reference [16] to calculate the density and viscosity of pure water, and the correctness of the results compared to the experimental results indicates that the code works properly for simulating the water.

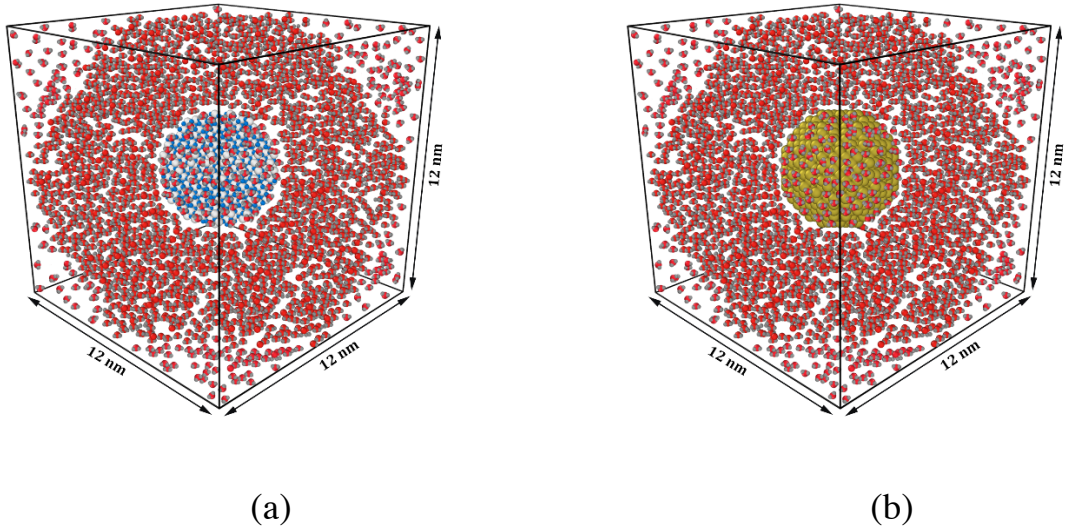

(a)          (b)

Fig. 1. Simulation box of (a) TiO$_2$-water (b) silver-water nanofluids (Ti blue, O white, Ow red, Hw gray, Ag yellow)

### 3- Theoretical background and nanofluid density calculation

To calculate the effective density of nanofluids the mixing theory has been commonly applied in most researches. This theory is based on the classical and statistical mechanics of ideal gas mixtures and does not include the geometry, particle dimension, nanoparticle distribution and ordering. This model expresses the nanofluid density as follows [13];

$$\rho_{nf-old} = \rho_{np}\varphi + \rho_{bf}(1-\varphi) \tag{5}$$

where *np* and *bf* denote the nanoparticle and base fluid respectively. Furthermore, ϕ is the nanoparticle volume fraction that can be obtained as:

$$\varphi = \frac{V_{np}}{V_{nf}} = \frac{V_{np}}{V_{np} + V_{bf,i}} \tag{6}$$

where $V_{bf,i}$ is the base fluid volume before mixing with nanoparticles.

The ternary mixture rule which was introduced previously in the ref [16] is employed to compute the density of nanofluid as follows:

$$\rho_{nf} = \varphi_{np}\rho_{np} + \varphi_{nl}\rho_{nl} + (1-\varphi_{nl}-\varphi_{np})\rho_{bf} \tag{7}$$

where

$$\varphi_{np} = \frac{V_{np}}{V_{np} + V_{bf,i} + V_{nl}\left(1-\frac{\rho_{nl}}{\rho_{bf}}\right)} \tag{8}$$

$$\varphi_{nl} = \frac{V_{nl}}{V_{np} + V_{bf,i} + V_{nl}\left(1-\frac{\rho_{nl}}{\rho_{bf}}\right)} \tag{9}$$

where $V_{nl}$ is the nanolayer volume. This volume for spherical nanoparticles is computed from Equation 10.

$$V_{nl} = \frac{4}{3}\pi\left(\left(\frac{d_{np}}{2}+t\right)^3 - \left(\frac{d_{np}}{2}\right)^3\right) \tag{10}$$

Moreover, the volume of spherical nanoparticle can be obtained from Equation 11 as follows:

$$V_{np} = \frac{4}{3}\pi\left(\frac{d_{np}}{2}\right)^3 \tag{11}$$

**4- Results and discussion**

**4.1 Effect of volume concentration**

The influence of different volume fractions on the density of nanofluid considering the contribution of formed nanolayer in the liquid/solid interface was examined. The question is whether the volume fraction has an influence on nanolayer properties such as nanolayer thickness as well as its density. To answer this question three nanofluids of silver-water with volume fractions of 2, 4 and 6% were simulated. Adding nanoparticle to the base fluid causes an inhomogeneous distribution of base fluid molecules throughout the simulation box which makes several local regions of density [33]. The volume around the nanoparticle in radial direction was divided into few spherical shells for computing the number density. The number density which is computed as follows [34] is used to describe the degree of concentration of water molecules in the simulation box.

$$n = \frac{\Delta N}{\Delta V} \tag{12}$$

where $n$ is the number density, $\Delta N$ is the quantity of water molecules and $\Delta V$ is the volume of spherical shell with the thickness of 0.1 nm around the nanoparticle.

The number density as a function of distance from the surface of silver nanoparticle with diameter of 4 nm suspended in water-based nanofluids with different volume concentrations are demonstrated in Fig 2. As it can be seen, the stronger potential of solid nanoparticle than liquid water has a great tendency to concentrate more water molecules in a closer volume of spherical shell. As the distance from the nanoparticle surface increases, the number density decreases and then it reaches to a constant value equal to the number density of the base fluid.

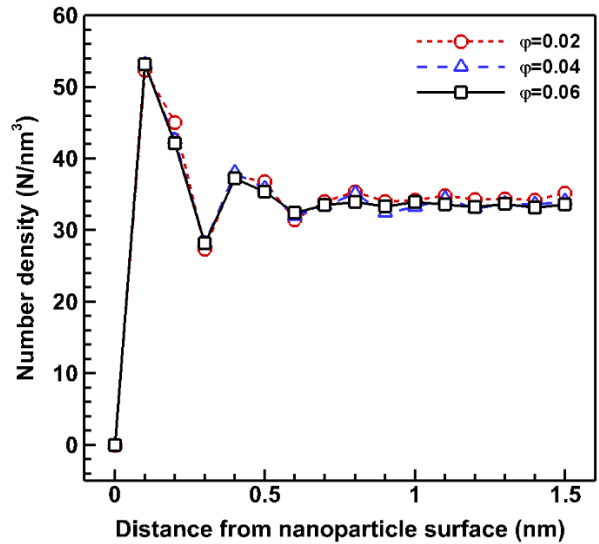

Fig. 2. Comparison of number density in absorption layers around the silver nanoparticle in silver-water nanofluids with different volume concentrations

Figure 3 compares the density fluctuations around the 4 nm diameter at three different volume concentrations. By adding more water to the vicinity of silver nanoparticle the governing physics of the problem remains unchanged and the density fluctuations would be naturally similar to the case in which the nominal volume fraction was 4%. The distance between the nanoparticle surface and the point that the density fluctuations are within a tolerance of 2% of the base fluid density is defined as the nanolayer thickness. Based on this criterion, the thicknesses of nanolayer in water-silver nanofluid are almost 0.9 nm which can be concluded that the volume fraction has no significant effect on density of nanolayer.

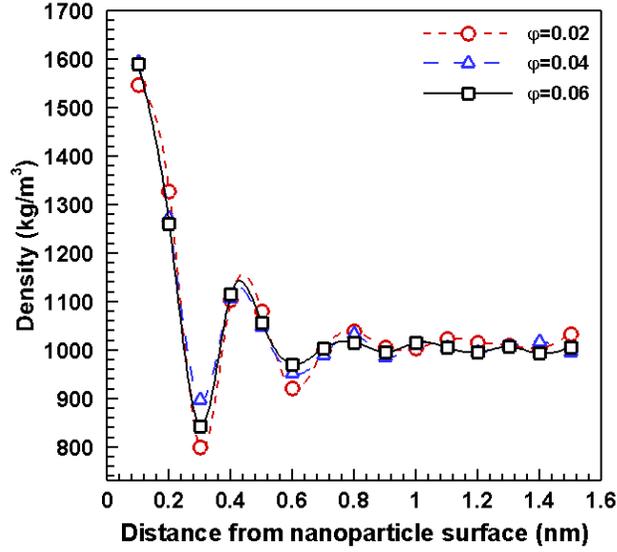

Fig. 3. Density fluctuations around the silver nanoparticle in water-based nanofluid with volume fractions of 2, 4 and 6%

Langmuir formula of monolayer adsorption of molecules, which is only dependent to the type of base fluid, confirms that monolayer is 0.28 nm in silver-water nanofluid (M = 18.015 g/mol and ϱ$_{bf}$ =0.997 g/cm³ [35]):

$$t = \frac{1}{\sqrt{3}} \left( \frac{4M_{bf}}{\rho_{bf} N_A} \right)^{\frac{1}{3}} \tag{13}$$

where $N_A$ is Avogadro's number (6.023 × 10²³ mol⁻¹) and $M_{bf}$ is the molecular weight of base fluid. Fig. 4 represents the microscopic schematic of nanolayer formation in water-based nanofluid.

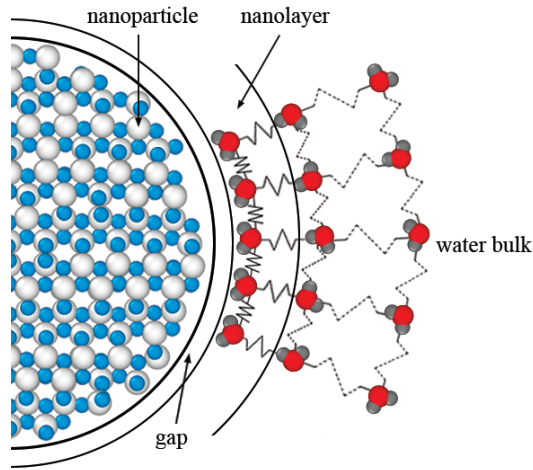

Fig. 4. The microscopic shematic of nanolayer formation around the TiO$_2$ nanoparticle in water-based nanofluid

Equation 14 is applied to calculate the density of nanolayer from MD outcomes [16]. Table 4 gives the best coefficients of equation 14 for calculating the density of nanolayer in silver-water nanofluid with different volume fractions as the nanolayer thickness is assumed to be 0.9 nm according to criterion of 2%.

$$\rho_{nl}(d_{np}) = a.e^{-\frac{b}{d_{np}^c}} \tag{14}$$

Table 4. The best curve fitted coefficients of Eq. 14 for silver-water nanofluids with different volume fraction; the nanolayer thickness is assumed to be 0.9 nm

| Coefficients of Eq. (14) | Volume fraction | | |
|---|---|---|---|
| | $\varphi = 0.02$ | $\varphi = 0.04$ [16] | $\varphi = 0.06$ |
| $a(kg/m^3)$ | 1081 | 1090 | 1078 |
| $c$ | 1.61 | 1.73 | 1.72 |
| $b(nm)$ | 0.5 | 0.5 | 0.5 |

Fig. 5 compares the outcomes of MD simulation, traditional binary mixture and the ternary mixture models for silver-water nanofluids with different volume fractions. As the volume concentration increases, the density of nanofluid is higher. But it should be noted that at low volume concentrations the nanolayer contribution to the density of nanofluid is small which will result in a smaller difference between the outcomes of ternary mixture model and traditional binary mixture one. This is exactly the same as Sharifpur et al. [15] empirically pointed out and attributed this to the influence of nanolayer on density of nanofluid.

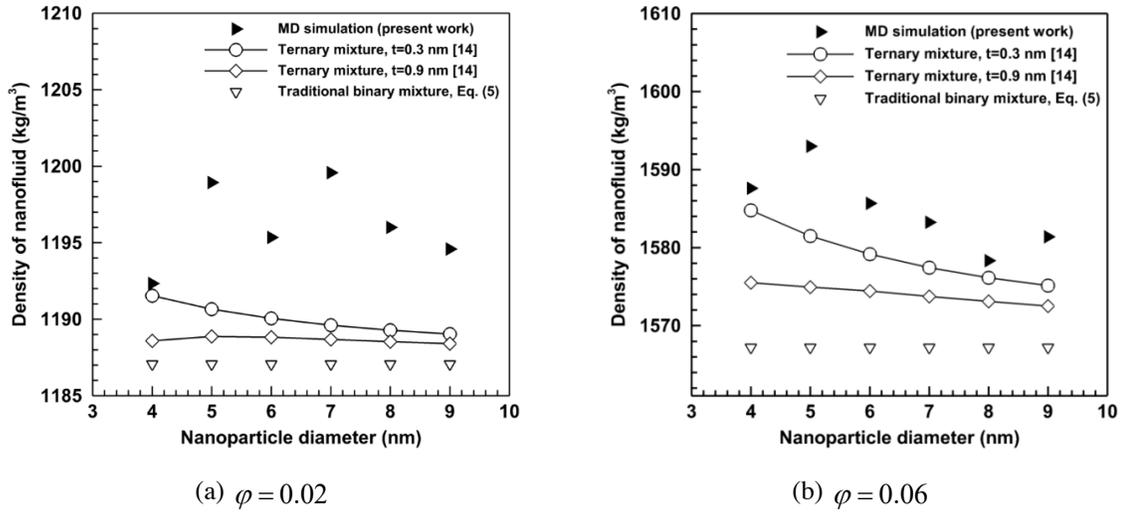

(a) $\varphi = 0.02$          (b) $\varphi = 0.06$

Fig. 5. Comparison of the density of silver nanofluids at volume concentrations of (a) 2% and (b) 6% for different values of nanolayer thicknesses in molecular dynamics simulation, traditional binary mixture and the ternary mixture models

Table 5 compares the nanofluid density values in three models of ternary mixture, traditional binary mixture, and Sharifpur et al. [13]. As it can be seen, the difference between the density of nanofluids in the ternary mixture model and the traditional binary mixture model grows with increasing nanoparticle concentration. Nevertheless, the model of Sharifpur et al. [13] shows lower nanofluid density values than the traditional binary mixture model while the density value in the ternary mixture model is higher than the traditional binary mixture model.

Table. 5. Nanofluid density values based on three different models

| Volume fraction, $\phi$ | Traditional binary mixture (kg/m$^3$) | Sharifpur model [15] (kg/m$^3$) | Ternary mixture (kg/m$^3$) |
| --- | --- | --- | --- |
| 0.02 | 1187.06 | 1185.64 | 1189.50 |
| 0.04 | 1377.12 | 1373.83 | 1380.38 |
| 0.06 | 1567.18 | 1561.58 | 1574.63 |

### 4.2 Effect of nanoparticle material

The number density along radial direction of the simulation box was calculated and obtained results are depicted in Figure 6. It is obvious that at close distances to the nanoparticle the number density for TiO$_2$ nanoparticle is smaller than silver nanoparticle. This can be due to the weak

Coulomb and LJ interactions between the charged atoms of $TiO_2$ nanoparticle and water molecules ($\varepsilon_{Ti-water}$ = 0.056 and $\varepsilon_{O-water}$ = 0.097 kcal/mol) compared to strong LJ interaction between Ag and water ($\varepsilon_{Ag-water}$ = 1.21 kcal/mol).

Figure 7 illustrates the density fluctuation around the silver and $TiO_2$ nanoparticles for nanoparticle diameter of 4 nm and volume fraction of 4%. As it can be seen, the form of density fluctuations around the $TiO_2$ nanoparticle is quite different from those around the silver nanoparticle. Density strongly fluctuates for $TiO_2$ nanoparticle at distances less than 0.5 nm. This is because, the potential energy of silver-water is stronger than $TiO_2$-water. Therefore, it is expected that the nanolayer density for silver nanofluid to be higher than that in $TiO_2$-water nanofluid. Based on the defined criterion of 2%, the thickness of nanolayer for $TiO_2$-water nanofluid is 0.5 nm while it is 0.9 nm for silver-water nanofluid.

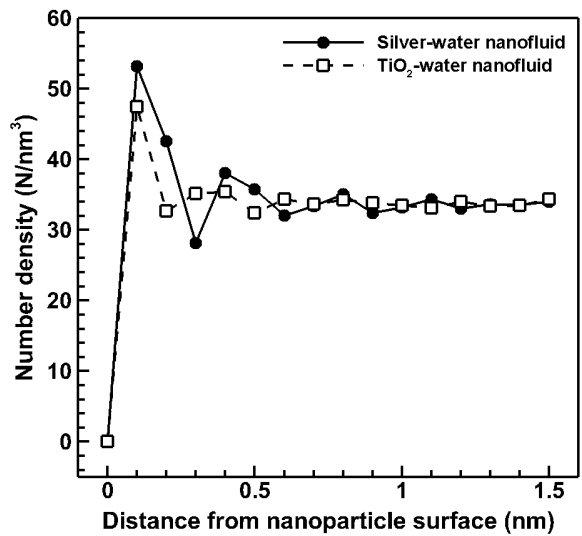

Fig. 6. Comparison of number densities around silver and $TiO_2$ nanoparticles

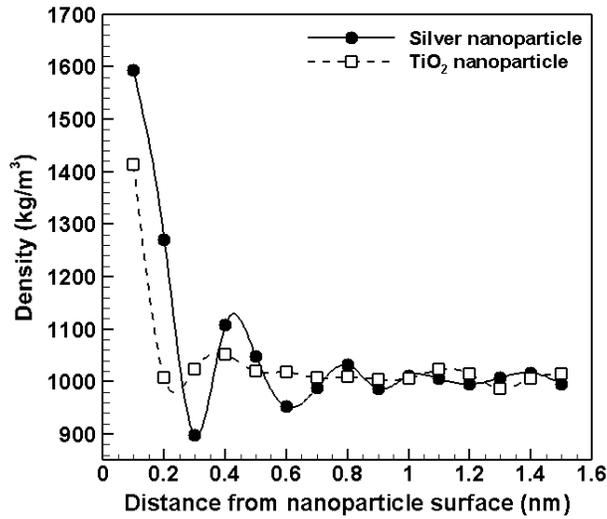

Fig. 7 The variation of water molecules density around the surface of silver and TiO$_2$ nanoparticles in water-based nanofluids at volume fraction of 4%

The best appropriate factors in Eq. 14 for estimating the density of nanolayer for TiO$_2$-water nanofluid are presented in Table 6.

Table 6. Best appropriate factors in Eq. 14 for estimating the density of nanolayer in TiO$_2$-water and silver-water nanofluids with 4% volume fraction

| Thickness of nanolayer (nm) | TiO$_2$-water nanofluid | | | silver-water nanofluid [16] | | |
|---|---|---|---|---|---|---|
| | $a(kg/m^3)$ | $c$ | $b(nm)$ | $a(kg/m^3)$ | $c$ | $b(nm)$ |
| 0.3 | 1280 | 1.05 | 0.5 | 1400 | 2.01 | 0.5 |
| 0.5 | 1169 | 1.19 | 0.5 | 1188 | 1.76 | 0.5 |

The hydrophobicity/hydrophilicity is a crucial surface property which determines the level of affinity of a surface with respect to water. The hydrophobicity is defined as the poor affinity of a solid surface toward water molecules and the hydrophilicity is defined as a strong affinity for water. TiO$_2$ due to its hydrophobic properties has a weak interaction with water molecules compared with silver nanoparticle (see Table 3). Hydrophobic nanoparticle makes a tiny gap of order of angstrom which is not observed around the silver nanoparticle. In contrast, silver as a hydrophilic nanoparticle has a tendency to attract nanolayer in a closer distance with higher density. Figure 8 illustrates the differences in the positioning of the water molecules about the TiO$_2$ and silver nanoparticles.

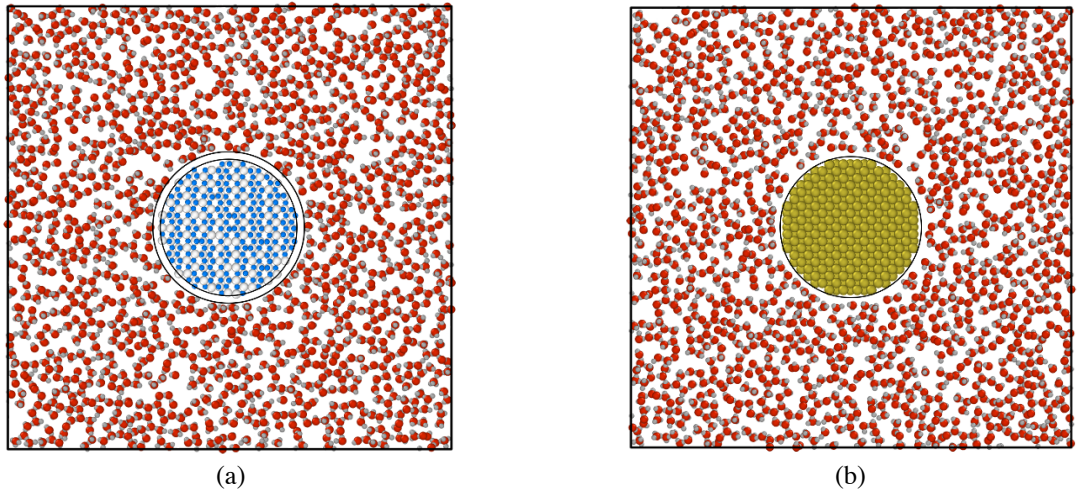

(a)                                        (b)

Fig. 8. (a) TiO$_2$ and (b) silver nanoparticle with diameter of 4 nm is surrounded by water molecules with volume fraction 2%. (Ti Blue, O White, Ow Red, Hw Gray)

The outcomes of three models, i.e. MD simulation, the ternary mixture and traditional binary mixture models, for calculating the density of nanofluids with different nanolayer thicknesses and different nanoparticle materials are compared in Fig 9. As the density of silver is greater than the density of TiO$_2$, the density of TiO$_2$-water nanofluid is lower than silver-water nanofluid. On the other hand, the value of nanofluid density in the ternary mixture model decreases with increasing the nanoparticle diameter. It can be interpreted as having large diameters, the $\varphi_{np,new}$ will be much larger than $\varphi_{nl,new}$, and the ternary mixture model will somehow turn into the binary mixture model.

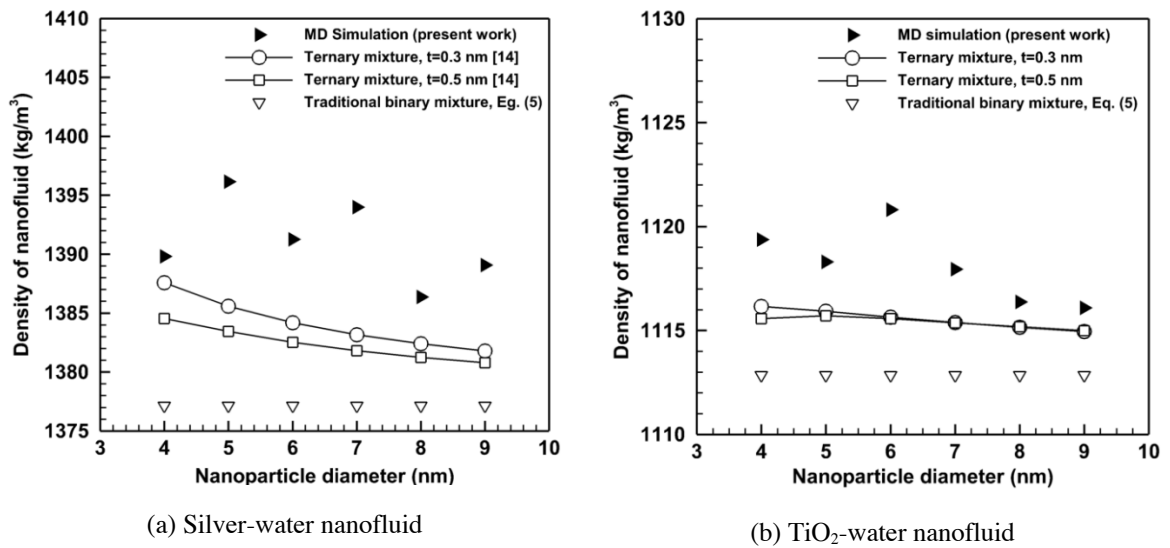

(a) Silver-water nanofluid                      (b) TiO$_2$-water nanofluid

Fig. 9. Comparison of the density of water-based nanofluids containing different nanoparticles (a) silver and (b) TiO$_2$ at volume fraction of 4%

## 5. Conclusion

The present study aimed to investigate the effect of type and volume fraction of nanoparticle on the density of nanofluid utilizing molecular dynamics simulations of silver and titanium dioxide water-based nanofluids. Silver (as a hydrophilic nanoparticle) and titanium dioxide (as a hydrophobic nanoparticle) were simulated by the well-known Lennard-Jones (LJ) potential and the Matsui-Akaogi force field, respectively. Outcomes revealed that the type of interatomic potential between nanoparticle and surrounding liquid can highly affect the thickness of nanolayer followed by density of nanolayer or nanofluid. The distance between the nanoparticle surface and the point that the density fluctuations are within a tolerance of 2% of the base fluid density is defined as the nanolayer thickness. Based on the criterion of 2%, the thickness of nanolayer was estimated 0.9 nm and 0.5 nm for silver-water nanofluid and $TiO_2$-water nanofluid, respectively. The density of nanolayer in silver-water nanofluid due to its strong solid-liquid interactions is larger than that of $TiO_2$-water nanofluid. It seems that the hydrophobicity/hydrophilicity of nanoparticle affect the density and thickness of nanolayer. Hydrophobic nanoparticle makes a tiny gap of order of an angstrom which is not observed around the hydrophilic nanoparticle. It was also found that the volume fraction of nanoparticle has no significant influence on the nanolayer thickness as well as the density of nanolayer. Moreover, the difference between the density of nanofluids in the ternary mixture model and the traditional binary mixture model increases as the volume fraction of nanofluids increases. This indicates that the effect of nanolayer gets more considerable at high volume fractions.

# 6. References

[1]  L. M. Corredor, M. M. Husein, and B. B. Maini, "Historical perspective A review of polymer nanohybrids for oil recovery," *Adv. Colloid Interface Sci.*, vol. 272, p. 102018, 2019, doi: 10.1016/j.cis.2019.102018.

[2]  H. Babar and H. M. Ali, "Towards hybrid nanofluids: preparation, thermophysical properties, applications, and challenges," *J. Mol. Liq.*, vol. 281, pp. 598–633, 2019.

[3]  A. Wahab, A. Hassan, M. A. Qasim, H. M. Ali, H. Babar, and M. U. Sajid, "Solar energy systems--Potential of nanofluids," *J. Mol. Liq.*, p. 111049, 2019.

[4]  L. Yang, J. Huang, W. Ji, and M. Mao, "Investigations of a new combined application of nanofluids in heat recovery and air purification," *Powder Technol.*, vol. 360, pp. 956–966, 2020.

[5]  A. J. Chamkha, M. Molana, A. Rahnama, and F. Ghadami, "On the nanofluids applications in microchannels: a comprehensive review," *Powder Technol.*, vol. 332, pp. 287–322, 2018.

[6]  S. A. Angayarkanni and J. Philip, "Review on thermal properties of nanofluids: recent developments," *Adv. Colloid Interface Sci.*, vol. 225, pp. 146–176, 2015.

[7]  L. Qiu *et al.*, "A review of recent advances in thermophysical properties at the nanoscale: From solid state to colloids," *Phys. Rep.*, vol. 843, pp. 1–81, 2020, doi: 10.1016/j.physrep.2019.12.001.

[8]  R. S. Vajjha, D. K. Das, and B. M. Mahagaonkar, "Density measurement of different nanofluids and their comparison with theory," *Pet. Sci. Technol.*, vol. 27, no. 6, pp. 612–624, 2009, doi: 10.1080/10916460701857714.

[9] T.-P. Teng and Y.-H. Hung, "Estimation and experimental study of the density and specific heat for alumina nanofluid," *J. Exp. Nanosci.*, vol. 9, no. 7, pp. 707–718, 2014.

[10] S. N. Shoghl, J. Jamali, and M. K. Moraveji, "Electrical conductivity, viscosity, and density of different nanofluids: An experimental study," *Exp. Therm. Fluid Sci.*, vol. 74, pp. 339–346, 2016.

[11] A. Naddaf and S. Z. Heris, "Density and rheological properties of different nanofluids based on diesel oil at different mass concentrations," *J. Therm. Anal. Calorim.*, vol. 135, no. 2, pp. 1229–1242, 2019.

[12] O. Mahian, A. Kianifar, and S. Wongwises, "Dispersion of ZnO nanoparticles in a mixture of ethylene glycol--water, exploration of temperature-dependent density, and sensitivity analysis," *J. Clust. Sci.*, vol. 24, no. 4, pp. 1103–1114, 2013.

[13] B. C. Pak and Y. I. Cho, "Hydrodynamic and heat transfer study of dispersed fluids with submicron metallic oxide particles," *Exp. Heat Transf. an Int. J.*, vol. 11, no. 2, pp. 151–170, 1998.

[14] O. Mahian *et al.*, "Recent advances in modeling and simulation of nanofluid flows-Part I: Fundamentals and theory," *Phys. Rep.*, vol. 790, pp. 1–48, 2019, doi: 10.1016/j.physrep.2018.11.004.

[15] M. Sharifpur, S. Yousefi, and J. P. Meyer, "A new model for density of nanofluids including nanolayer," *Int. Commun. Heat Mass Transf.*, vol. 78, pp. 168–174, Nov. 2016, doi: 10.1016/j.icheatmasstransfer.2016.09.010.

[16] M. M. Heyhat, A. Rajabpour, M. Abbasi, and S. Arabha, "Importance of nanolayer formation in nanofluid properties: Equilibrium molecular dynamics simulations for Ag-

water nanofluid," *J. Mol. Liq.*, vol. 264, pp. 699–705, 2018, doi: 10.1016/j.molliq.2018.05.122.

[17] M. Abbasi, M. M. Heyhat, and A. Rajabpour, "Study of the effects of particle shape and base fluid type on density of nanofluids using ternary mixture formula: A molecular dynamics simulation," *J. Mol. Liq.*, vol. 305, p. 112831, 2020, doi: 10.1016/j.molliq.2020.112831.

[18] B. Mortazavi, E. V. Podryabinkin, S. Roche, T. Rabczuk, X. Zhuang, and A. V. Shapeev, "Machine-learning interatomic potentials enable first-principles multiscale modeling of lattice thermal conductivity in graphene/borophene heterostructures," *Mater. Horizons*, vol. 7, no. 9, pp. 2359–2367, 2020, doi: 10.1039/d0mh00787k.

[19] J. L. F. Abascal and C. Vega, "A general purpose model for the condensed phases of water: TIP4P/2005," *J. Chem. Phys.*, vol. 123, no. 23, p. 234505, 2005.

[20] H. Zhang and J. F. Banfield, "Thermodynamic analysis of phase stability of nanocrystalline titania," *J. Mater. Chem.*, vol. 8, no. 9, pp. 2073–2076, 1998.

[21] "Materials Studio Materials Modeling & Simulation Application | Dassault Systèmes BIOVIA." .

[22] R. W. G. Wyckoff, *Crystal structures*. Krieger, 1964.

[23] P. K. Naicker, P. T. Cummings, H. Zhang, and J. F. Banfield, "Characterization of titanium dioxide nanoparticles using molecular dynamics simulations," *J. Phys. Chem. B*, vol. 109, no. 32, pp. 15243–15249, 2005, doi: 10.1021/jp050963q.

[24] V. N. Koparde and P. T. Cummings, "Molecular dynamics simulation of titanium dioxide

nanoparticle sintering," *J. Phys. Chem. B*, vol. 109, no. 51, pp. 24280–24287, 2005, doi: 10.1021/jp054667p.

[25] V. N. Koparde and P. T. Cummings, "Molecular dynamics study of water adsorption on TiO2 nanoparticles," *J. Phys. Chem. C*, vol. 111, no. 19, pp. 6920–6926, 2007, doi: 10.1021/jp0666380.

[26] B. Luan, T. Huynh, and R. Zhou, "Simplified TiO2force fields for studies of its interaction with biomolecules," *J. Chem. Phys.*, vol. 142, no. 23, p. 234102, 2015, doi: 10.1063/1.4922618.

[27] M. Matsui and M. Akaogi, "Molecular dynamics simulation of the structural and physical properties of the four polymorphs of TiO2," *Mol. Simul.*, vol. 6, no. 4–6, pp. 239–244, 1991, doi: 10.1080/08927029108022432.

[28] "Titanium Dioxide - Titania ( TiO2)." .

[29] J. H. Park and N. R. Aluru, "Temperature-dependent wettability on a titanium dioxide surface," *Mol. Simul.*, vol. 35, no. 1–2, pp. 31–37, 2009, doi: 10.1080/08927020802398884.

[30] P. Guan, D. R. Mckenzie, and B. A. Pailthorpe, "MD simulations of Ag film growth using the Lennard-Jones potential," vol. 8753, 1996, doi: 10.1016/j.socscimed.2010.02.039.

[31] S. Plimpton, "Fast parallel algorithms for short-range molecular dynamics," *J. Comput. Phys.*, vol. 117, no. 1, pp. 1–19, 1995, [Online]. Available: http://lammps.sandia.gov.

[32] L. Martinez, R. Andrade, E. G. Birgin, and J. M. Martinez, "PACKMOL: a package for building initial configurations for molecular dynamics simulations," *J. Comput. Chem.*, vol. 30, no. 13, pp. 2157–2164, 2009, doi: 10.1002/jcc.

[33] W. Cui, Z. Shen, J. Yang, and S. Wu, "Molecular dynamics simulation on the microstructure of absorption layer at the liquid-solid interface in nanofluids," *Int. Commun. Heat Mass Transf.*, vol. 71, pp. 75–85, 2016, doi: 10.1016/j.icheatmasstransfer.2015.12.023.

[34] L. Li, Y. Zhang, H. Ma, M. Yang, Æ. Y. Zhang, and Æ. H. Ma, "Molecular dynamics simulation of effect of liquid layering around the nanoparticle on the enhanced thermal conductivity of nanofluids," *J. Nanoparticle Res.*, vol. 12, no. 3, pp. 811–821, 2010, doi: 10.1007/s11051-009-9728-5.

[35] R. E. Sonntag, C. Borgnakke, G. J. Van Wylen, and S. Van Wyk, *Fundamentals of thermodynamics*, vol. 6. Wiley New York, 1998.